\documentclass[%
 reprint,
nofootinbib,
 amsmath,amssymb,
 aps,
]{revtex4-2}

\usepackage{graphicx}
\usepackage{dcolumn}
\usepackage{bm}
\usepackage{amsmath}
\usepackage{caption} 
\usepackage{subcaption}
\usepackage{amssymb}
\usepackage{pifont}
\RequirePackage{textpos}
\usepackage[symbol]{footmisc}

\skip\footins 0.5cm

\begin{document}

\preprint{APS/123-QED}

\title{Probing CPV mixing in the Higgs sector in VBF at 1 TeV ILC}

\author{N. Vuka\u{s}inovi\'{c}}
\email{nvukasinovic@vin.bg.ac.rs}
 
\author{I. Bo\v{z}ovi\'{c}-Jelisav\u{c}i\'{c}}
\author{G. Ka\u{c}arevi\'{c}}
\author{I. Smiljani\'{c}}%
\author{I. Vidakovi\'{c}} 
\affiliation{``VIN\u{C}A'' Institute of Nuclear Sciences - National Institute of the Republic of Serbia, University of Belgrade, 11001 Belgrade, Serbia} 

\date{\today}


\begin{abstract}
With the current precision of measurements by the ATLAS and CMS experiments, it cannot be excluded that a SM-like Higgs boson is a CP violating mixture of CP-even and CP-odd states. We explore this possibility here, assuming Higgs boson production in ZZ-fusion, at 1 TeV ILC, with unpolarized beams.  The full simulation of SM background and fast simulation of the signal is performed, simulating 8 ab$^{-1}$ of data collected with the ILD detector. We demonstrate that the CP mixing angle $\Psi_{\mathrm{CP}}$ between scalar and pseudoscalar states can be measured with the statistical uncertainty of 3.8 mrad at 68\% CL, corresponding to 1.44 $\cdot 10^{-5}$ for the CP parameter $f_\mathrm{CP}$, for the pure scalar state.  This is the first result on sensitivity of an $e^{+}e^{-}$ collider to measure $f_\mathrm{CP}$ in the Higgs production vertex in vector boson fusion.      
\end{abstract}

\maketitle


\section{\label{sec:intro} Introduction} 

Since the experimentally established CP violation (CPV) in the quark sector is not sufficient to explain the baryon asymmetry of the observable Universe, exploring the possibility that CP is violated in the Higgs sector is an important part of the physics program both at ongoing  experiments and future Higgs factories. Although a purely CP-odd state for the Higgs boson is already excluded by the ATLAS and CMS experiments \cite{ratlas, rcms}, there is still a possibility that CP symmetry is violated in Higgs interactions with bosons and fermions.

Experiments at the International Linear Collider (ILC) \cite{ilc1, ilc2} will be able to explore a plethora of Higgs production and decay mechanisms to probe bosonic and fermionic vertices for CPV. A tentative list of the processes of interest at ILC is illustrated in Table \ref{table:1} \cite{rtabela1}. The CPV effect is typically weaker in Higgs interactions with vector bosons ($HVV$) in comparison to those with fermions ($Hff$) since the pseudoscalar state does not directly couple to the Standard Model (SM) particles, and sensitivity targets to measure CPV effects in these interactions are thus different. In order to provide a common platform for interpretation of the CPV measurements in bosonic $HVV$ and fermionic $Hff$ vertices, as well as to interpret projections for different future experiments, a common framework is defined in \cite{rsnowm}, based on the CPV parameter $f_\mathrm{CP}$. The parameter $f_\mathrm{CP}$ quantifies the CP-odd contribution to a Higgs interaction \cite{rsnowm} as:

\begin{equation}
\label{fcp}
f_{CP}^{hX} = \frac{\Gamma_{h\rightarrow X}^{CP^{\mathrm{odd}}}}{\Gamma_{h\rightarrow X}^{CP^{\mathrm{odd}}} + \Gamma_{h\rightarrow X}^{CP^{\mathrm{even}}}}
\end{equation}

\noindent assuming here the Higgs decay to a final state $X$. In order to explain the baryon asymmetry of the Universe assuming the 2HDM model of an extend Higgs sector, a minimal 10\% contribution of the CP-odd state is required \cite{rsnowm}. This sets the theoretical target for future colliders to measure $f_\mathrm{CP}$ with an absolute precision better than 10$^{-2}$ (10$^{-6}$) in $Hff$ ($HVV$) vertices. The state-of-the-art projections on $f_\mathrm{CP}$ sensitivity at different colliders to measure a pure scalar state with 68\% CL are given in Table \ref{table:2}. The projected precision of measurements in $HVV (V = Z, W)$ vertices at future $e^{-}e^{+}$ colliders is based on \cite{rgritsan} where the estimate is given for the $HZZ$ production vertex in Higgstrahlung, at 250 GeV center-of-mass energy, assuming 2.5 ab$^{-1}$ of data. The study is performed at the generator level and thus without realistic simulation of a detector response for signal and background.  Estimates at higher center-of-mass energies for $HVV$ production vertices (marked with `*' in Table \ref{table:2}) are obtained by scaling of the result at 250 GeV to the corresponding integrated luminosities indicated in Table \ref{table:2}.

\begin{table}[tb]
\centering
\caption{\label{table:1} Possible Higgs production and decay modes to probe CPV at various center-of-mass energies at ILC.}
\begin{tabular}{ |p{2.7cm}|p{2.7cm}p{2.7cm}| } 
 \hline
 \hline
 {mode} & \multicolumn{2}{c|} {fermion couplings} \\
 \hline
 decay & H$\rightarrow \tau^{+}\tau^{-}$ & 250+ GeV \\ 
 production & $e^{-}e^{+}\rightarrow Ht\bar{t}$ & 500+ GeV \\
 \hline
 {} & \multicolumn{2}{c|} {boson couplings} \\
 \hline
 production & $e^{-}e^{+}\rightarrow HZ$ & 250+ GeV \\ 
 decay & H$\rightarrow ZZ$ & 250+ GeV \\ 
 decay & H$\rightarrow WW$ & 250+ GeV \\ 
 production & $e^{-}e^{+}\rightarrow He^{-}e^{+}$ & 1000+ GeV \\ 
 \hline
 \hline
\end{tabular}
\end{table}

\begin{figure}[tb]
\centering
\includegraphics[width=0.7\columnwidth]{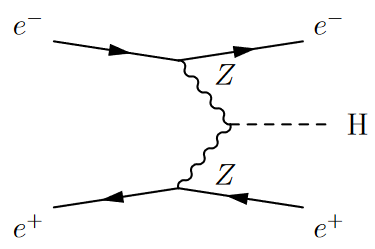}
\caption{\label{fig-zz} Feynman diagram of the signal process.} 
\end{figure}

\begin{table*}[tb]
\centering
\caption{\label{table:2} Expected precision of the CP parameter $f_{CP}$ (68\% CL) for measurements in $HXX$ ($X = f,V$) vertices for the pure scalar state at various colliders. The mark `$\checkmark$' indicates feasibility of such a measurement. The entries marked with `*' are obtained by scaling the precision to measure $f_{CP}$ in $HZZ$ interaction at 250 GeV to the corresponding integrated luminosities indicated in the table. This table is adapted from Table I of \cite{rsnowm}. }

\begin{tabular}{ |p{2cm}|p{1.3cm} p{1.3cm} p{1cm} p{1.5cm} p{1.7cm} p{1.7cm} p{1.7cm} p{0.7cm} p{0.7cm} p{0.7cm} p{0.7cm}|p{1.05cm}| } 
\hline
\hline
Collider & $pp$ & $pp$ & $pp$ & $e^{-}e^{+}$ & $e^{-}e^{+}$ & $e^{-}e^{+}$ & $e^{-}e^{+}$ & $e^{-}p$ & $\gamma\gamma$ & $\mu^{+}\mu^{-}$ & $\mu^{+}\mu^{-}$ & target\\
E(GeV) & 14,000 & 14,000 & 100,000 & 250 & 350 & 500 & 1,000 & 1,300 & 125 & 125 & 3,000 & (theory) \\
$\mathcal{L}$ (fb$^{-1}$) & 300 & 3,000 & 30,000 & 2500 & 3500 & 5000 & 10,000 & 1,000 & 250 & 20 & 1,000 & \\
\hline
\hline
$HZZ/HWW$ & 4.0 $\cdot 10^{-5} $ & 2.5 $\cdot 10^{-6}$ & \checkmark & 3.9 $\cdot {10^{-5}}$  & 2.9 $\cdot {10^{-5}}^{(*)}$ & 1.3 $\cdot {10^{-5}}^{(*)}$ & 3.0 $\cdot {10^{-6}}^{(*)}$ & \checkmark & \checkmark & \checkmark & \checkmark & $< 10^{-5}$ \\
\hline
\hline
$\mathcal{L}$ (fb$^{-1}$) & 300 & 3,000 & 30,000 & 250 & 350 & 500 & 1,000 & 1,000 & 250 & 20 & 1,000 &  \\ \hline \hline
$H\gamma\gamma$ &  & 0.50 & \checkmark &  & & & & & 0.06 & & & $< 10^{-2}$ \\ \hline
$HZ\gamma$ &  & $\sim$ 1 & \checkmark & & &  & $\sim$ 1 & & & & & $< 10^{-2}$ \\ \hline
$Hgg$ & 0.12 & 0.011 & \checkmark &  & & & & & & & & $< 10^{-2}$ \\
\hline
\hline
$Ht\bar{t}$ & 0.24 & 0.05 & \checkmark & & & 0.29 & 0.08 & \checkmark &  & & \checkmark & $< 10^{-2}$ \\ \hline 
$H\tau\tau$ & 0.07  & 0.008 & \checkmark & 0.01 & 0.01 & 0.02 & 0.06 &  & \checkmark & \checkmark & \checkmark & $< 10^{-2}$ \\ \hline
$H\mu\mu$ &  &  &  &  & &  &  &   &  & \checkmark &  & $< 10^{-2}$ \\
\hline
\hline
\end{tabular}
\end{table*}

The analysis presented below is the first result obtained for the Higgs production in vector boson fusion (VBF), specifically in $ZZ$-fusion $e^{-}e^{+}\rightarrow He^{-}e^{+}$, assuming 8 ab$^{-1}$ of data collected with the ILC operating at 1 TeV center-of-mass-energy with unpolarised beams. Since $ZZ$-fusion is a t-channel process (Fig. \ref{fig-zz}), electrons and positrons in signal events are peaked at small polar angles. The interplay between the production cross-section and centrality of signal events makes 1 TeV an optimal energy for CPV studies of the $HZZ$ vertex in VBF at an $e^{+}e^{-}$ collider. 

The paper is organized as follows: Section \ref{sec:sec2} introduces event samples and software tools, methodology of the measurement is discussed in Section \ref{sec:sec3A}, while statistical interpretation and discussion of the obtained results are given in Section \ref{sec:sec3B}.

\section{\label{sec:sec2}Event samples}
In this analysis we consider Higgs boson production in $ZZ$-fusion (Fig. \ref{fig-zz}) with the Higgs boson decaying to $b\bar{b}$ with a branching fraction of $\sim$ 56\% in the SM \cite{r1}. Exclusive reconstruction of the Higgs decays enables us to avoid the high cross-section $e^{-}e^{+} \rightarrow e^{-}e^{+}\gamma$ background that would otherwise be present in an inclusive analysis. 
We have generated 6 $\cdot 10^{5}$ signal events in Whizard 2.8.3 \cite{rwhiz}, using the Higgs characterization model \cite{rhc} within the UFO framework to allow contribution of the CP-odd component to the 125 GeV Higgs mass eigenstate. In this model the parametrization of  CP mixing is entirely realized in terms of the mixing angle between scalar and pseudoscalar states, allowing for a completely general description of CP-mixed states \cite{rhc}\footnote{In order to remove total cross section dependence on the mixing angle, copuling of the CP-odd component in the effective Lagrangian of \cite{rhc} was set to $\kappa_{AZZ}\,=\,3.57$, for the CP-even couplings $\kappa_{HZZ} = 0$ and $\kappa_{H\partial Z} = 1$.}. 

Further interactions of signal with the detector are simulated assuming a generic detector for ILC with the fast simulation DELPHES 3.4.2 (ILCgen cards) \cite{rdelphes}. For a smaller sample of signal events ($\sim$ 3500 events) the response of the ILD detector \cite{rild} is fully simulated with detailed Geant4 \cite{rgeant4} simulation implemented in the Mokka toolkit \cite{rmokka}. These events are reconstructed using realistic Marlin processors \cite{rmarlin}. The Standard Model background is also fully simulated and reconstructed with the ILD detector, using the same simulation tools as for the signal. Backgrounds due to beamstrahlung and hadron photo-production are overlaid onto the fully simulated events in the digitization phase.
Fragmentation and hadronization are simulated in Pythia 6.4 \cite{rphytia} for all events. Particle identification is based on the Particle Flow Approach implemented in the DDMarlinPandora processor \cite{rddmarlin} of Marlin and, in a simplified version, in the fast simulation DELPHES. 
The considered processes and their cross-sections are given in Table \ref{table:3}.

\begin{table}[tb]
\centering
\caption{\label{table:3} Cross-sections for signal and background processes with expected number of events in the full range of polar angles in 8 $\mathrm{ab^{-1}}$ of data, and number of reconstructed Monte Carlo events.}
\renewcommand*{\thempfootnote}{\fnsymbol{mpfootnote}}
\begin{tabular}{ |p{2.5cm}|p{1.3cm} p{1.8cm} p{2.4cm}| } 
\hline
\hline
Signal     & {$\sigma (fb) $}  & {Expected \hspace{1cm} in 8 $\mathrm{ab^{-1}}$} & {Reconstructed \hspace{1cm} MC events} \\
\hline 
$e^-e^+\rightarrow He^-e^+$; $H\rightarrow b \bar{b}$      & 13	& 104000	& \begin{tabular}{l} 6 $\cdot 10^5$\footnote[2]{DELPHES} \\3495\footnote[8]{full sim.} \end{tabular}  \\
\hline
\multicolumn{4}{|l|}{Background$^{\dag\dag}$} \\
\hline 
$e^-e^+\rightarrow q\bar{q}e^+e^-$    & 2.4 $\cdot 10^3$	& 19 $\cdot 10^6$ & 2 $\cdot 10^5$	 \\
\hline
$e^-e^+\rightarrow qq$             & 3.6 $\cdot 10^3$	 & 29 $\cdot 10^6$	& 4 $\cdot 10^5$    \\
\hline
$e^-e^+\rightarrow q\bar{q}e\nu$    & 3 $\cdot 10^3$		& 24 $\cdot 10^6$	& 2.6 $\cdot 10^6$  \\
\hline
$e^-e^+\rightarrow llll$    & 8 $\cdot 10^3$		& 64 $\cdot 10^6$	& 1.5 $\cdot 10^6$   \\
\hline
$e^-e^+\rightarrow eeqqqq$    & 37	& 30 $\cdot 10^4$ & 1 $\cdot 10^4$  \\
\hline
$e^-e^+\rightarrow e\nu_{e}qqqq$    & 51	& 4 $\cdot 10^5$	& 1 $\cdot 10^6$  \\ 
\hline
$e^-e^+\rightarrow qq\nu_{e}ee\nu_{e}$    &  5.6	& 45 $\cdot 10^3$ 	& 5 $\cdot 10^4$   \\
\hline
\hline
\end{tabular}
\end{table}

Since the signal signature is one electron-positron pair and two $b$-jets in the final state, the event selection is based on identification of exactly one isolated electron and one isolated positron per event while the remaining Particle Flow Objects (PFOs) are clustered into two jets by the Durham algorithm \cite{rdurham}. Electron and positron candidates are required each to have energy above 60\,GeV. 

Electron isolation is based on different observables for events processed in fast or full detector simulation. For signal events processed in DELPHES, electrons are considered isolated if there is no additional particle with transverse momentum greater than 0.5 GeV in a radius $R = $ 0.5 cone in $\eta-\phi$ space around the electron candidate direction, and if the sum of transverse momenta of all other particles within the cone is less than 12\% of the transverse momentum of the electron candidate. For fully simulated and reconstructed events, electrons are selected if their transverse and longitudinal impact parameters are less than 0.1 mm and 1 mm respectively, ratio of depositions in electromagnetic versus hadronic calorimeter is larger than 0.95, and the additional energy in a cone of size 0.1 rad around the electron direction satisfies $E_{\mathrm cone}^{\mathrm{2}} < 40 \, \mathrm{GeV} \cdot E_{e^{\pm}} - 20\, \mathrm{GeV^2}$.

A Multivariate Analysis (MVA) is used to further reduce the contribution from several high cross-section background processes, in particular $e^-e^+\rightarrow q\bar{q}e^+e^-$ with its signal-like signature. The Boosted Decision Tree (BDT) classifier is employed, as implemented in the TMVA toolkit \cite{rmva}. Ten input observables are used: di-jet invariant mass $m_{j\bar{j}}$, invariant mass $m_{e^+e^-}$ and transverse momentum $p_T{_{ee}}$ of the final state $e^+e^-$ system, polar angle of the di-jet system $\theta_{j\bar{j}}$, number of particle flow objects per event $N_{\mathrm{PFO}}$, energies of final state $e^+$ and $e^-$, transverse momenta of jets $p_T{_{j_1}}$ and $p_T{_{j_2}}$. Additional requirements are applied to suppress background after the MVA application: $m_{j\bar{j}}> 110$ GeV, $p_T{_{j_2}}> 160$ GeV and $N_{\mathrm{PFO}_{1,2}}> 10$, where $N_{\mathrm{PFO}_{1, 2}}$ stands for the number of particle flow objects in jets.   
The signal efficiency is obtained as the ratio of the number of selected signal events and the number of signal events with both $e^+$ and $e^-$ in the tracking region of polar angles ($|\cos\theta| < $ 0.98) and it is found to be 70\%.  Only two MC background events remain after the selection, corresponding to around 240 background events expected in 8\,ab$^{-1}$ of data.
Selected signal and background are collectively referred to as reconstructed data in further text, if not stated otherwise. 

\section{\label{sec:sec3}Method}
\begin{figure}[tb]
\centering
\includegraphics[width=0.8\columnwidth]{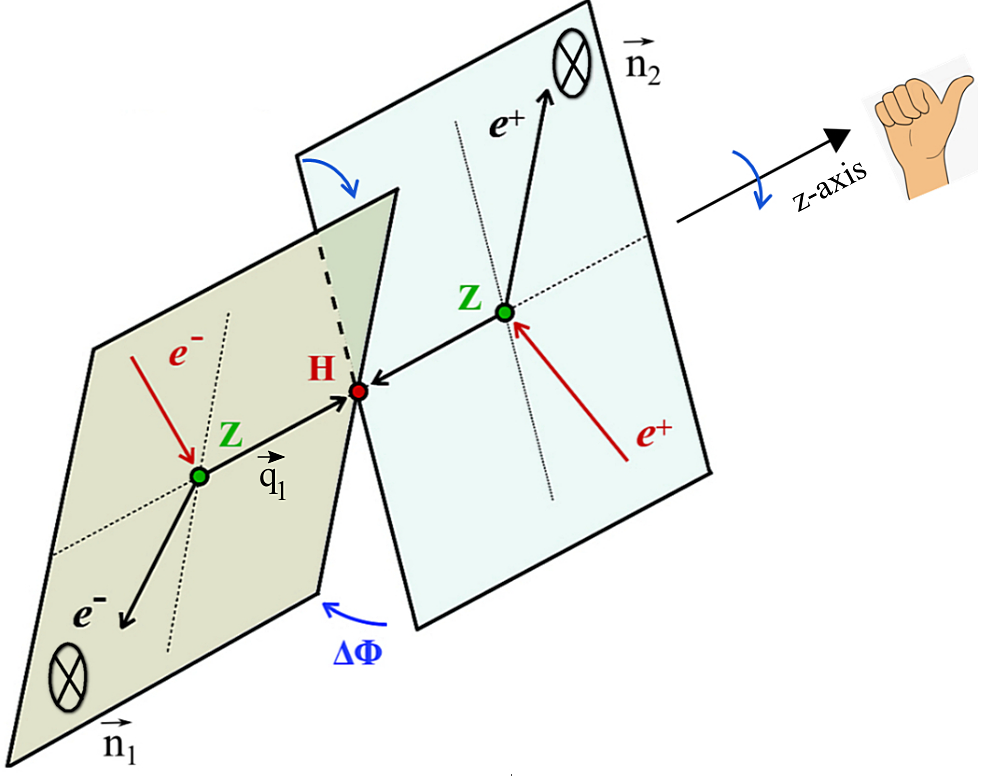}
\caption{\label{fig-delphi} Illustration of $\Delta\Phi$ angle between the scattering planes, defined in the Higgs boson rest frame. Initial electron and positron are drawn in red. }
\end{figure}

In the Standard Model the Higgs boson is a CP-even state with the CP-conserving couplings. In models with an extended Higgs sector, the 125 GeV Higgs mass eigenstate ($h$) could be a mixture of CP-even ($H$) and CP-odd ($A$) states:

\begin{equation}
\label{hmix}
h = H \cdot \cos\Psi_{\mathrm{CP}} + A \cdot \sin\Psi_{\mathrm{CP}},
\end{equation}
\noindent where $\Psi_{\mathrm{CP}}$ is the mixing angle violating CP symmetry in Higgs interactions for non-zero values. There are several observables sensitive to non-zero values of $\Psi_{\mathrm{CP}}$ \cite{rogawa}, one of the most sensitive is the angle $\Delta\Phi$ between scattering planes illustrated in Fig. \ref{fig-delphi}. $\Delta\Phi$ is the rotation angle of the positron plane with respect to the electron plane in the Higgs reference frame, around the axis defined by the momentum of the $Z$ boson emitted by the initial electron, following the right-hand rule. 

\begin{figure}[tb]
\centering
\includegraphics[width=0.9\columnwidth]{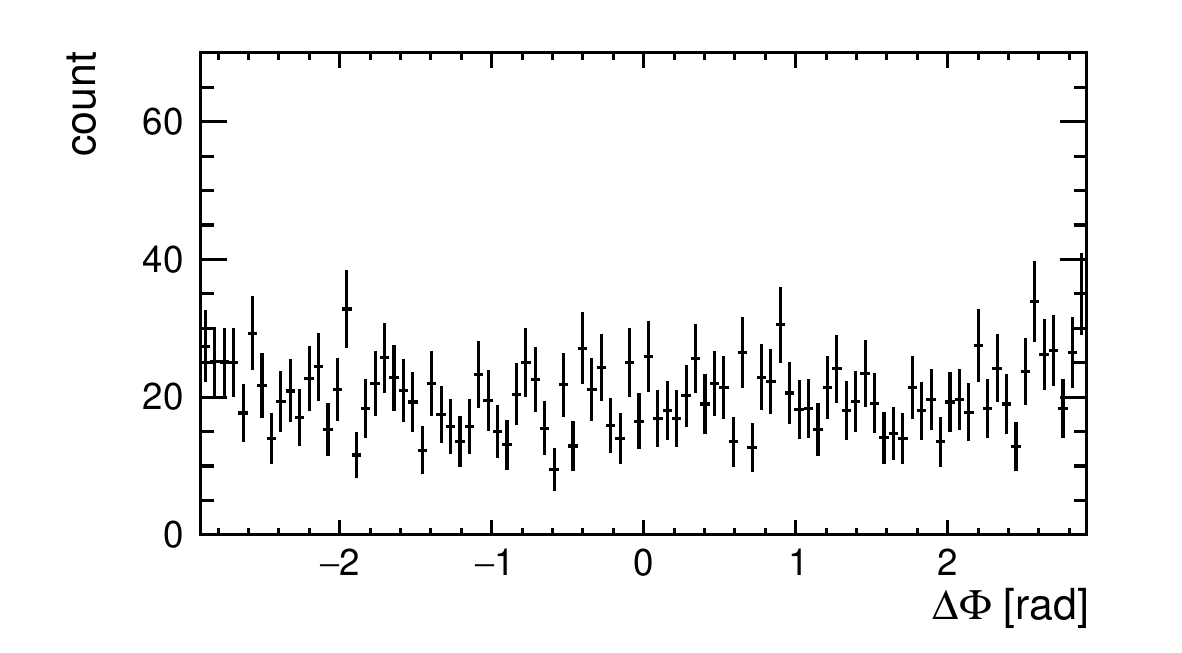}
\begin{textblock}{0.08}(4.9,-2.05) 
\textbf{ILD}
\end{textblock}
\caption{\label{fig-fi_qqll} $\Delta\Phi$ distribution of $e^-e^+\rightarrow q\bar{q}e^+e^-$ background after preselection of one isolated $e^-$ and $e^+$ per event, with energies greater than 60 GeV. } 
\end{figure}

$\Delta\Phi$ can be calculated as the angle between unit vectors ($\overrightarrow{n}_{1}$ and $\overrightarrow{n}_{2}$) orthogonal to electron and positron scattering planes, respectively:
\begin{equation}
\label{fi} 
\Delta\Phi = 
\mathrm{sgn}(\Delta\Phi) \cdot \arccos(\overrightarrow{n}_{1}\cdot\overrightarrow{n}_{2})  
\end{equation}
\noindent where:

\begin{equation}
\label{eqsin}
\mathrm{sgn}(\Delta\Phi) = \frac{ \overrightarrow{q}_{1}\cdot (\overrightarrow{n}_{1}\times\overrightarrow{n}_{2}) }{ |\overrightarrow{q}_{1}\cdot (\overrightarrow{n}_{1}\times\overrightarrow{n}_{2})|},
\end{equation}

and:
\begin{equation}
\label{n12}
\overrightarrow{n}_{1} = \frac{ \overrightarrow{q}_{e^-_{i}}\times \overrightarrow{q}_{e^-_{f}} }{ |\overrightarrow{q}_{e^-_{i}}\times \overrightarrow{q}_{e^-_{f}}| } \hspace{.3cm} \mathrm{and} \hspace{.3cm} \overrightarrow{n}_{2} = \frac{ \overrightarrow{q}_{e^+_{i}}\times \overrightarrow{q}_{e^+_{f}} }{ |\overrightarrow{q}_{e^+_{i}}\times \overrightarrow{q}_{e^+_{f}}|},
\end{equation}
\noindent $\overrightarrow{q}_{e^{-(+)}_{i(f)}}$ is the momentum of initial (final) state electron (positron) and $\overrightarrow{q}_{1}$ is momentum of the $Z$ boson emitted by the initial electron. 

\begin{figure}[tb]
\centering 
\includegraphics[width=0.9\columnwidth]{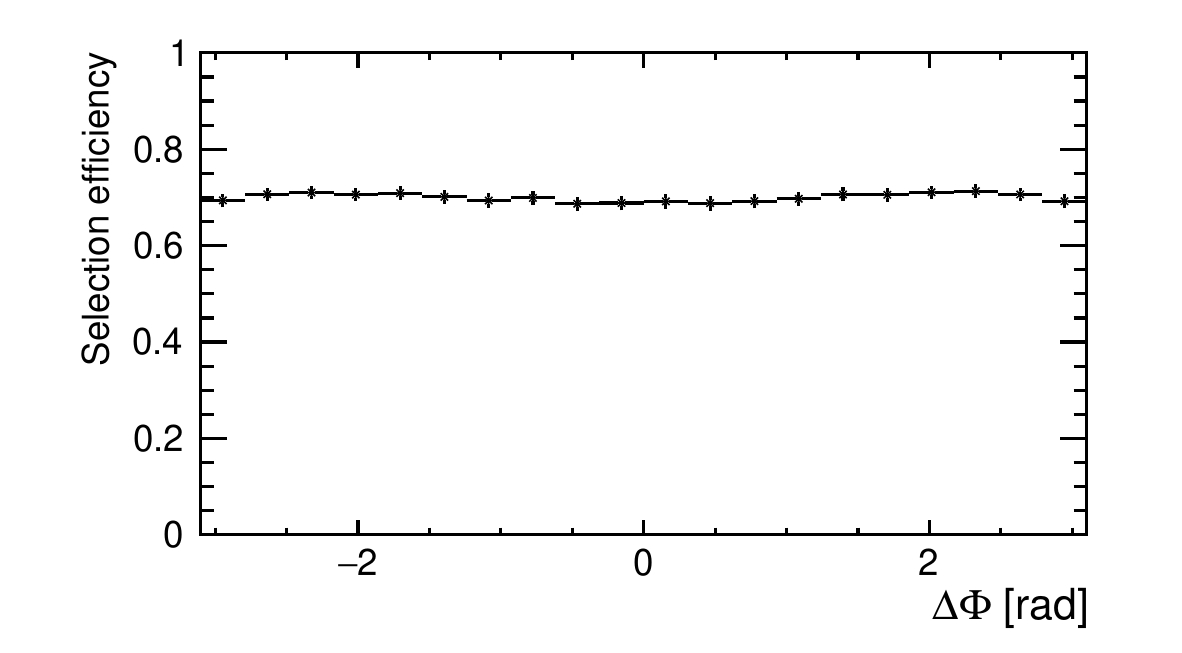}
\caption{\label{fig-seleff} Signal selection efficiency w.r.t. the sensitive observable $\Delta\Phi$.}
\end{figure}

The distribution of $\Delta\Phi$ for background is flat reflecting the fact that background is CP insensitive, as illustrated for $q\bar{q}e^+e^-$ final state background remaining after selection of one isolated $e^-$ and $e^+$ per event with energies greater than 60 GeV (Fig. \ref{fig-fi_qqll}). The event selection described in the previous section does not bias the sensitive observable. This is illustrated in Fig. \ref{fig-seleff}.

\begin{figure}[tb]
\centering
\includegraphics[width=\columnwidth]{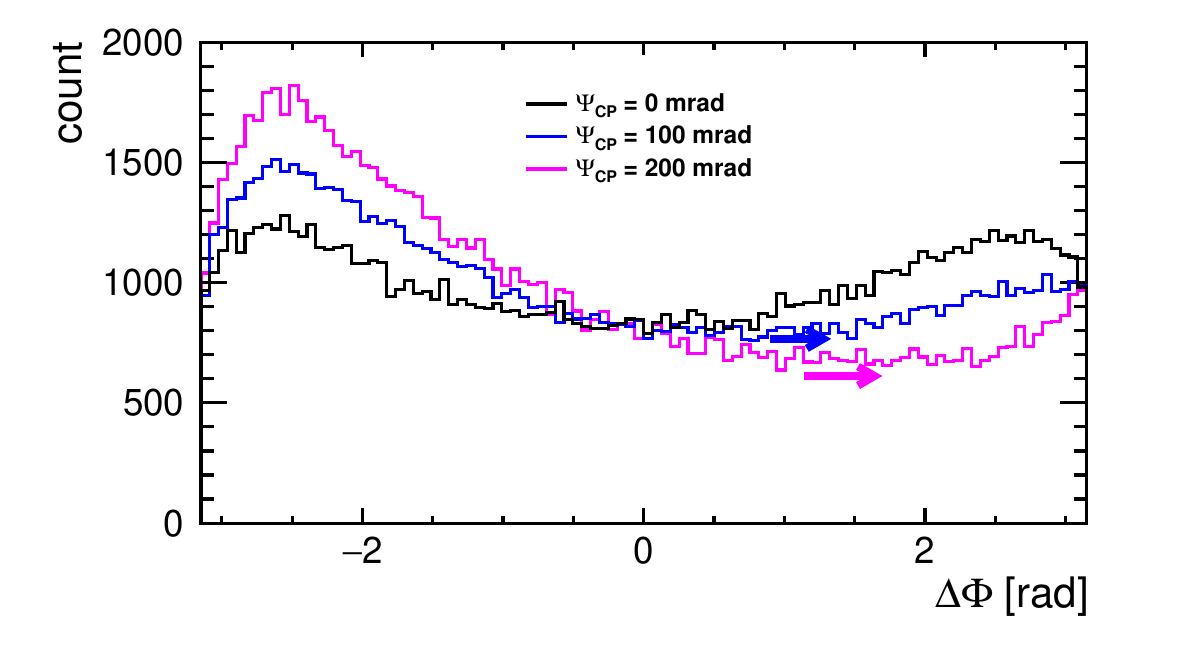}
\caption{\label{fig-shift_fi} $\Delta\Phi$ distribution for different $\Psi_{\mathrm{CP}}$ values illustrating the shift of the $\Delta\Phi$ minimum for non-zero values of $\Psi_{\mathrm{CP}}$.} 
\end{figure}

\begin{figure}[tb]
\centering
\includegraphics[width=\columnwidth]{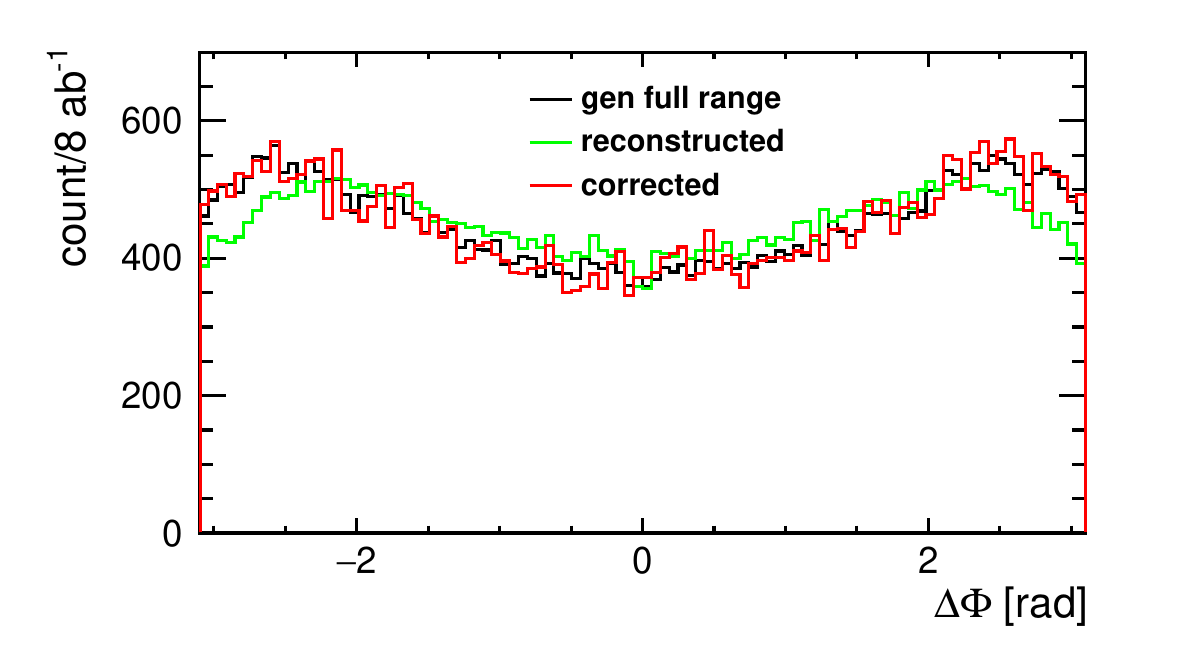} 
\caption{\label{fig-gen_rec_corr} $\Delta\Phi$ distributions for generated signal in the full physical range of polar angles, reconstructed signal in DELPHES and reconstructed signal corrected for the limited detector acceptance. } 
\end{figure}

\begin{figure}[tb]
\centering
\includegraphics[width=\columnwidth]{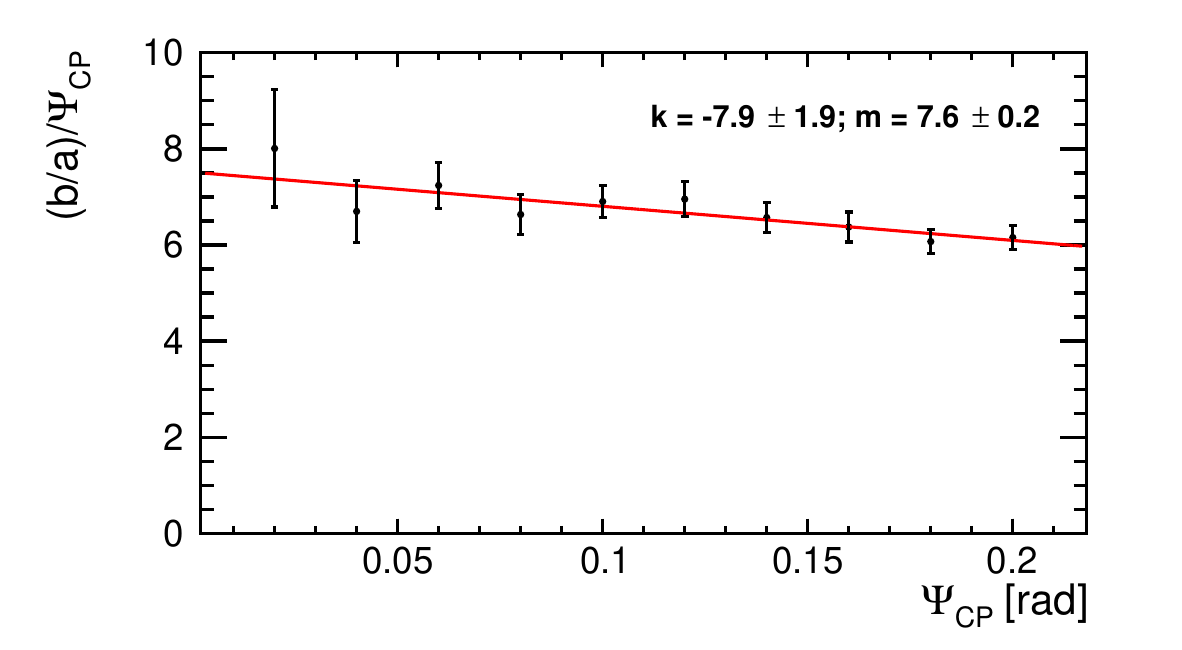} 
\caption{\label{fig-k_m} Positions of the minima of $\Delta\Phi$ distributions ($b/a$) over the true values of $\Psi_{\mathrm{CP}}$. }
\end{figure}
\subsection{\label{sec:sec3A} $\Delta\Phi$ and CP mixing angle $\Psi_{\mathrm{CP}}$}
Differently from $Hf\bar{f}$ vertices where the dependence of $\Delta\Phi$ on $\Psi_{\mathrm{CP}}$ can be derived from the differential cross-section for $H\rightarrow f\bar{f}$ decay \cite{rjeans}, CP violating contributions in bosonic $HVV$ vertices occur at the loop level and there is no simple analytical dependence of the sensitive observable $\Delta\Phi$ on the CP mixing angle $\Psi_{\mathrm{CP}}$. The dependence is therefore to be empirically determined, in this case by correlating the position of the minimum of $\Delta\Phi$ distribution to the true value of the mixing angle $\Psi_{\mathrm{CP}}$ used in event generation. As can be seen from Fig. \ref{fig-shift_fi}, the position of the minimum of $\Delta\Phi$ shifts to larger values for positive values of $\Psi_{\mathrm{CP}}$ (and similarly to the left for negative values of $\Psi_{\mathrm{CP}}$). 
Before determining the position of the minimum, reconstructed data has to be corrected for effects of detector acceptance, in order to retrieve the information on the CP state of the Higgs boson in the full physical range of polar angles. Fig.\,\ref{fig-gen_rec_corr} illustrates the $\Delta\Phi$ distribution from generated signal in the full physical range, reconstructed signal with the fast simulation and the signal corrected for the detector acceptance to compensate for the limited acceptance in polar angles of the central detector tracking system. The acceptance function is obtained as the ratio of the generated $\Delta\Phi$ distributions for signal in the central tracking region ($|\cos\theta| <$ 0.98) and in the full range of polar angles.  

\begin{figure}[tb]
\centering
\includegraphics[width=\columnwidth]{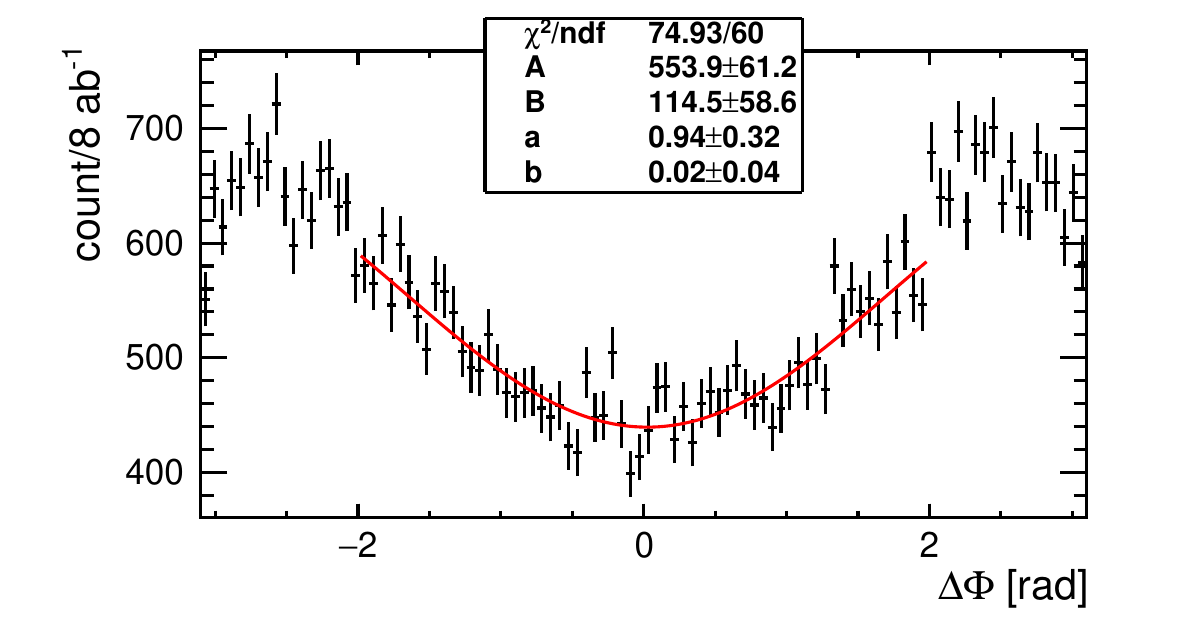}  
\caption{\label{fig-reco} Fit of $\Delta\Phi$ distribution of the selected and corrected reconstructed data with the function $f(\Delta\Phi)$ from Eq. \ref{fitf}, in order to obtain the position of the minimum ($b/a$). } 
\end{figure}

\begin{figure}[tb]
\centering
\includegraphics[width=\columnwidth]{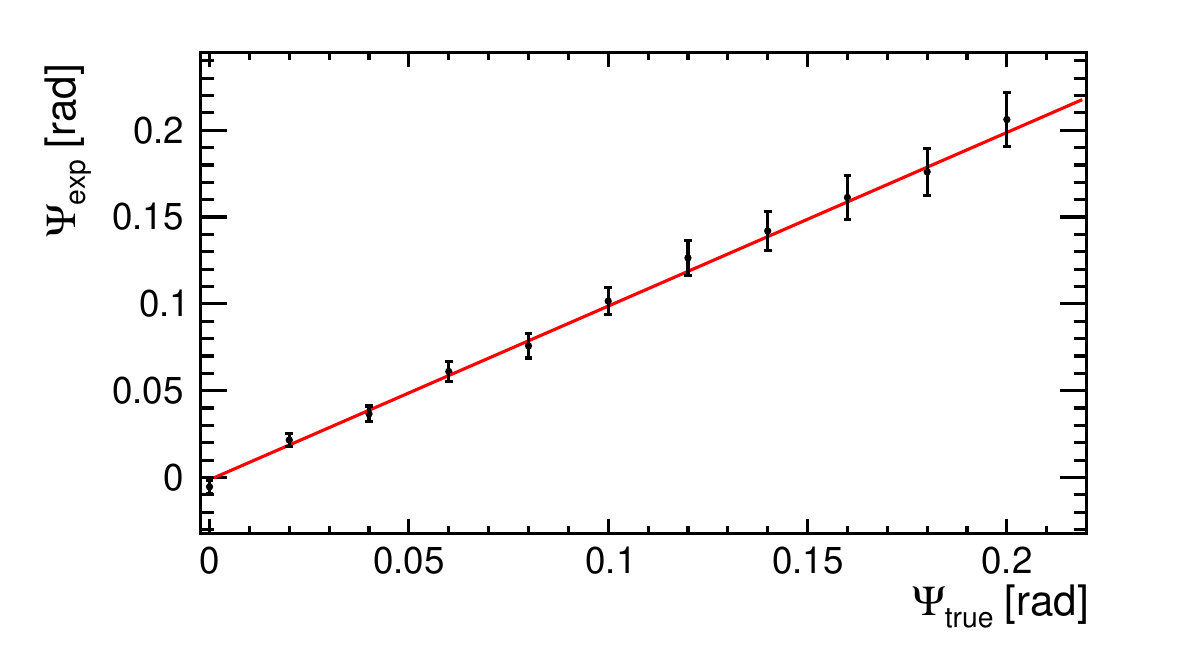}
\caption{\label{fig-true_exp} $\Psi_{\mathrm{CP}}$ values ($\Psi_{\mathrm{exp}}$) determined from Eq. \ref{quadf} versus $\Psi_{\mathrm{CP}}$ true values ($\Psi_{\mathrm{true}}$). The red line is illustrating the ideal dependence $\Psi_{\mathrm{exp}} = \Psi_{\mathrm{true}}$.}
\end{figure}
The minimum of $\Delta\Phi$ distribution from the reconstructed data can be determined by a local fit with the function $f(\Delta\Phi)$:
\begin{equation}
\label{fitf}
f(\Delta\Phi) = A + B \cdot \cos(a \cdot \Delta\Phi - b)
\end{equation}
\noindent where $A, B, a$ and $b$ are free parameters. From the principle of the first derivative, the ratio $b/a$ determines the minimum of $\Delta\Phi$ distribution.

\begin{figure}[tb]  
\begin{subfigure} {\columnwidth}
\caption{}\label{fig:11a}
\centering
\includegraphics[width=\columnwidth]{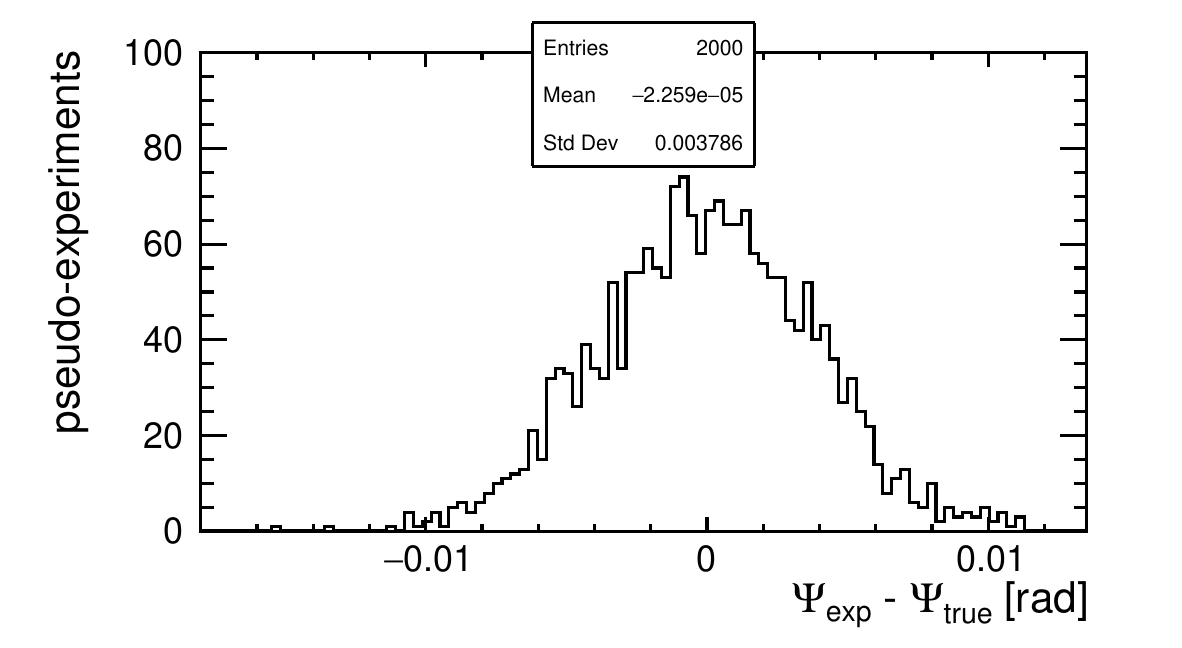}
 \end{subfigure}
\quad
\begin{subfigure}{\columnwidth}  
\caption{}\label{fig:11b}
\includegraphics[width=\columnwidth]{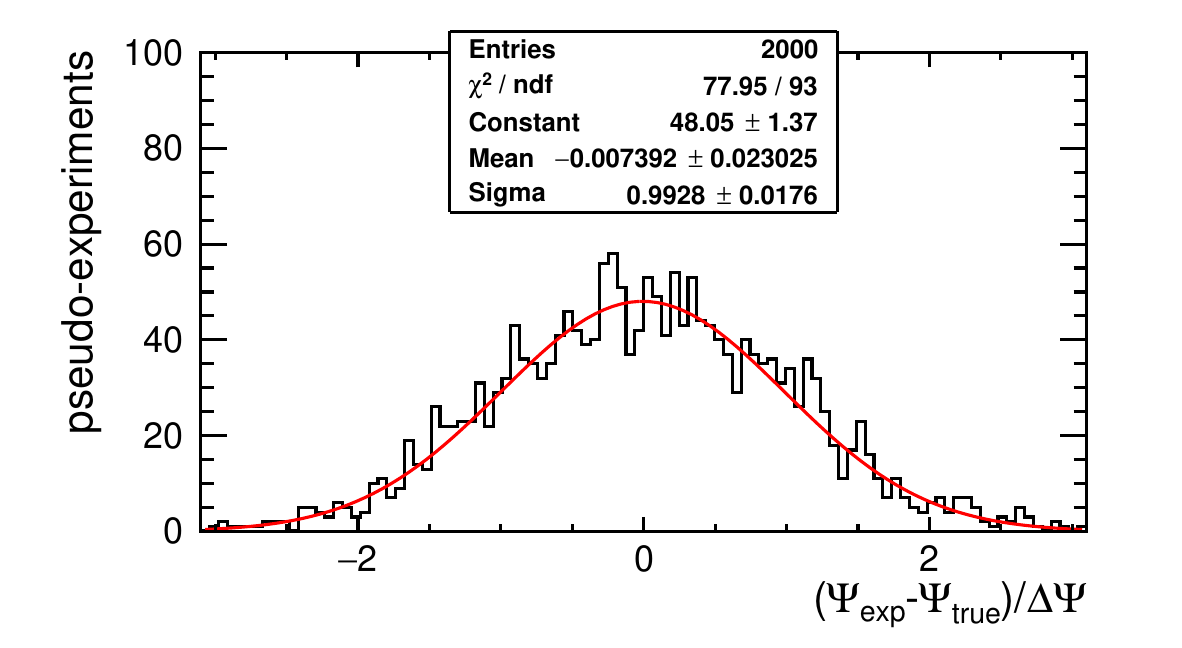}     
\end{subfigure}
\caption{\label{fig:11} (a) Statistical dissipation of measured $\Psi_{\mathrm{CP}}$ values ($\Psi_{\mathrm{exp}}$) w.r.t. the true ones ($\Psi_{\mathrm{true}}$). (b) Pull distribution for 2000 pseudo-experiments.}
\end{figure}

For $\Psi_{\mathrm{CP}}$ values up to 200 mrad, the variable $(b/a)/\Psi_{\mathrm{CP}}$ is to a good approximation a linear function of true values of $\Psi_{\mathrm{CP}}$, as shown in Fig. \ref{fig-k_m}, with coefficients $k$ and $m$ determining a slope and a constant term, respectively. Knowing the parameters $k$ and $m$ from simulation, $\Psi_{\mathrm{CP}}$ values can be determined by solving the quadratic equation:

\begin{equation}
\label{quadf}
k \cdot \Psi^{2}_{\mathrm{CP}} + m \cdot \Psi_{\mathrm{CP}} - (b/a) = 0
\end{equation}
\noindent where the minimum $b/a$ is measured from experimental or in this case from the reconstructed pseudo-data. The fit of reconstructed data corrected for the detector acceptance is illustrated in Fig. \ref{fig-reco}. The Fig. \ref{fig-true_exp} illustrates that $\Psi_{\mathrm{CP}}$ values extracted this way are in agreement with the true ones within the statistical uncertainties. Statistical uncertainties are derived from the uncertainties of the fit parameters $a$ and $b$ from Eq. \ref{fitf} and from the uncertainties of parameters $k$ and $m$ from Eq. \ref{quadf}. The method is applicable to measure CP mixing angles up to approximately 200 mrad above which the $\chi^2$ fit with the function $f(\Delta\Phi)$ (Eq. \ref{fitf}) significantly deteriorates.  
\subsection{\label{sec:sec3B} Statistical uncertainty and interpretation of the measurement}
From the fit to the single pseudo-experiment assuming 8 ab$^{-1}$ of data illustrated in Fig. \ref{fig-reco}, one determines $\Psi_{\mathrm{CP}}$\,= (2.4 $\pm$ 4.0) mrad by solving Eq. \ref{quadf}. In order to estimate the statistical dispersion of results of repeated $\Psi_{\mathrm{CP}}$ measurements, we performed 2000 pseudo-experiments each with 8 ab$^{-1}$ of data. The dispersion of the results assuming a pure scalar state is found to be 3.8 mrad at 68\% CL, as illustrated in Fig. \ref{fig:11} (a). Dispersion of errors from repeated pseudoexperiments is 0.4 mrad. The pull distribution shown in Fig. \ref{fig:11} (b) illustrates that the estimate of the statistical uncertainty on $\Psi_{\mathrm{CP}}$ is reasonable. Allowing parameters $k$ and $m$ (from Eq. \ref{quadf}) to vary within their uncertainties, we have estimated a systematic uncertainty from modeling to be significantly less than 1\,mrad. 

To interpret the obtained precision of measurement of the mixing angle in terms of sensitivity to the CP-odd amplitude $f_\mathrm{CP}^{HZZ}$,  following \cite{rsnowm} we  assume that $f_\mathrm{CP}^{HZZ}$ will vary from zero as $\sin^{2}(\Delta(\Psi_{\mathrm{CP}}))$ for the pure scalar state, where 
$\Delta(\Psi_{\mathrm{CP}})$ is the absolute statistical uncertainty of the $\Psi_{\mathrm{CP}}$ measurement. The statistical uncertainty of 3.8\,mrad of the $\Psi_{\mathrm{CP}}$ determination translates into $f_\mathrm{CP}^{HZZ}$ sensitivity of 1.44 $\cdot 10^{-5}$ at 68\% CL. The comparable results can be obtained if polarized samples of signal and background are considered, as the ILC operation foresees 80\% (20\%) polarization for electron (positron) beams at 1 TeV center-of-mass energy.

\section{\label{sec:sec4}Conclusion}
This analysis brings the first result of the CP mixing angle measurement in $HVV$ interactions where the Higgs boson is produced in vector boson fusion. We assume $Hee$ production in $ZZ - $ fusion with the reconstruction of exclusive Higgs decays to $b\bar{b}$ with the fast detector simulation. Standard Model background is fully simulated assuming the ILD detector response to 8\,ab$^{-1}$ of data collected at 1 TeV center-of-mass energy with unpolarised beams.
This measurement relies on the model-independent hypothesis that the 125\,GeV Higgs mass eigenstate could be a mixture of CP-even and CP-odd states with the mixing angle $\Psi_{\mathrm{CP}}$. From the shape of distribution of the CP sensitive angle between the scattering planes, the mixing angle $\Psi_{\mathrm{CP}}$ can be extracted with a statistical uncertainty of 3.8 mrad at 68\% CL, for the pure scalar state. This translates to the sensitivity of the CP parameter $f_\mathrm{CP}^{HZZ}$ of 1.44 $\cdot 10^{-5}$.  

\begin{acknowledgments}
This research is institutionally funded by the Ministry of Science and Technological Development of the Republic of Serbia and by the Science Fund of the Republic of Serbia through the Grant No. 7699827, IDEAS HIGHTONE-P.
Authors would also like to acknowledge our colleagues from ILC IDT Working Group 3 and the ILD Detector Concept Group for useful discussions, and especially to Prof. Aleksander Filip \.{Z}arnecki for sharing ideas leading to a better understanding of the sensitive observable behavior w.r.t. the CP mixing angle. We are also grateful to Dr. Daniel Jeans for careful reading of the text.  
\end{acknowledgments}


\begin{thebibliography}{99}
\bibitem{ratlas} ATLAS Collaboration, \textit{Measurement of the Higgs boson coupling properties in the $H\rightarrow ZZ^\ast \rightarrow 4l$ decay channel at $\sqrt{s} = $13 TeV with the ATLAS detector}, \href{http://dx.doi.org/10.1007/JHEP03(2018)095}{J. High Energy Phys. 03, 095} (2018).

\bibitem{rcms} CMS Collaboration, \textit{Study of the Mass and Spin-Parity of the Higgs Boson Candidate Via Its Decays to Z Boson Pairs}, \href{https://journals.aps.org/prl/abstract/10.1103/PhysRevLett.110.081803}{Phys. Rev. Lett. 110, 081803} (2013).

\bibitem{ilc1} C. Adolphsen et al., \textit{The International Linear Collider Technical Design Report - Volume 3.II: Accelerator Baseline Design},  ILC-REPORT-2013-040, \href{https://arxiv.org/ftp/arxiv/papers/1306/1306.6328.pdf}{arXiv:1306.6328} [physics.acc-ph] (2013).


\bibitem{ilc2} A. Aryshev, T. Behnke et al., \textit{The International Linear Collider: Report to Snowmass 2021}, DESY-22-045, \href{https://arxiv.org/pdf/2203.07622.pdf}{arXiv:2203.07622v3} [physics.acc-ph] (2023).


\bibitem{rtabela1}
D. Jeans et al., \emph{Measuring the CP properties of the Higgs sector at electron-positron colliders}, Letter of Interest for SnowMass2021: Energy Frontier (2020).

\bibitem{rsnowm}
A. V. Gritsan et. al, \emph{Snowmass White Paper: Prospects of CP-violation measurements with the Higgs boson at future experiments}, \href{https://arxiv.org/pdf/2205.07715.pdf}{ arXiv:2205.07715v2 [hep-ex]} (2022).

\bibitem{rgritsan}
I. Anderson, S. Bolognesi et. al, \emph{Constraining anomalous HVV interactions at proton and lepton colliders}, 
\href{ https://journals.aps.org/prd/abstract/10.1103/PhysRevD.89.035007}{Phys. Rev. D 89, 035007} (2014).

\bibitem{r1} S. Dittmaier et al., \textit{Handbook of LHC Higgs Cross Sections: 2. Differential Distributions}, Report No. CERN-2012-002 (2012). 


\bibitem{rwhiz} W. Kilian, T. Ohl, and J. Reuter, \textit{WHIZARD: Simulating multi-particle processes at LHC and ILC}, \href{https://doi.org/10.1140/epjc/s10052-011-1742-y}{Eur. Phys. J. C 71, 1742 (2011).}

\bibitem{rhc}
P. Artoisenet et al., \emph{A framework for Higgs characterization}, Journal of High Energy Physics 11, 043 (2013).
\bibitem{rdelphes}
J. de Favereau et al., \emph{DELPHES 3: a modular framework for fast simulation of a generic collider experiment}, Journal of High Energy Physics 2014, 57, \href{https://arxiv.org/abs/1307.6346}{arXiv:1307.6346} [hep-ex] (2014). 

\bibitem{rild}
H. Abramowicz et al. (ILD Concept Group), \emph{International Large Detector: Interim Design Report}, DESY-20-034, \href{https://doi.org/10.48550/arXiv.2003.01116}{arXiv:2003.01116 } [physics.ins-det] (2020).

\bibitem{rgeant4}
S. Agostinelli \textit{et al.}, \emph{Geant4 - A Simulation Toolkit}, \href{https://doi.org/10.1016/S0168-9002(03)01368-8}{Nucl. Instrum. Methods Phys. Res., Sect. A 506, 250 (2003).}

\bibitem{rmokka}
P. Mora de Freitas and H. Videau, \emph{Detector Simulation with Mokka/Geant4: Present and Future}, International Workshop on Linear Colliders, JeJu Island, Korea, Technical Report No. LC-TOOL-2003-010, 2002.

\bibitem{rmarlin} F. Gaede, \emph{Marlin and LCCD — Software tools for the ILC} \href{https://doi.org/10.1016/j.nima.2005.11.138}{Nucl. Instrum. Methods A 559, 177 (2006).}

\bibitem{rphytia}
T. Sjostrand, S. Mrenna, and P. Z. Skands, \emph{PYTHIA 6.4 Physics and Manual}, \href{https://doi.org/10.1088/1126-6708/2006/05/026}{J. High Energy Phys. 05 (2006) 026.}

\bibitem{rddmarlin}
O. Wendt, F. Gaede, and T. Kramer, \emph{Event reconstruction with MarlinReco at the ILC}, \href{https://doi.org/10.1007/s12043-007-0237-8}{Pramana 69 1109 (2007). }


\bibitem{rdurham}
S. Catani, Y. L. Dokshitzer, M. Olsson, G. Turnock, and B. Webber, \emph{New clustering algorithm for multi-jet cross-sections in $e^+e^-$ annihilation}, Phys.Lett. B269 (1991).

\bibitem{rmva}
A. H\"{o}cker \textit{et al.}, TMVA - Toolkit for multivariate data analysis, \href{http://arxiv.org/abs/physics/0703039}{arXiv:physics/0703039.}


\bibitem{rogawa}
T. Ogawa, \emph{Sensitivity to anomalous VVH couplings induced by dimension-6 operators at the ILC}, PhD thesis, Hayama, Japan (2018).

\bibitem{rjeans}
D. Jeans and G. W. Wilson, \emph{Measuring the CP state of tau lepton pairs from Higgs decay at the ILC}, Phys. Rev. D 98 013007 (2018).



\end{thebibliography}
\end{document}